\documentstyle[aaspp4,epsfig]{article}

\slugcomment{OU-TAP 96; ApJL, in press}

\begin{document}

\title{The X-ray Fundamental Plane and $L_{\rm X}-T$ Relation of
  Clusters of Galaxies} 
\author{Yutaka Fujita and Fumio Takahara}
\affil{Department of Earth and Space Science, Graduate School of
  Science, Osaka University, Machikaneyama-cho, Toyonaka, Osaka,
  560-0043, Japan} \authoremail{fujita@vega.ess.sci.osaka-u.ac.jp}

\begin{abstract}
  We analyze the relations among central gas density, core radius, and
  temperature of X-ray clusters by plotting the observational data in
  the three-dimensional ($\log \rho_0$, $\log R$, and $\log T$) space
  and find that the data lie on a 'fundamental plane'. Its existence
  implies that the clusters form a two-parameter family. The data on the
  plane still has a correlation and form a band on the plane. The
  observed relation $L_{\rm X} \propto T^3$ turns out to be the cross
  section of the band perpendicular to the major axis, while the major
  axis is found to describe the virial density. We discuss implications
  of this two-parameter family nature of X-ray clusters.
  
\end{abstract}


\section{Introduction}
\label{sec:intro}

Correlations among physical quantities of clusters of galaxies are
very useful tools for studying formation of clusters and cosmological
parameters. In particular, the luminosity ($L_{\rm X}$)-temperature ($T$) 
relation in X-ray clusters has been studied by many authors.
Observations show that clusters of galaxies exhibit a correlation of
approximately $L_{\rm X} \propto T^3$ (Edge \& Stewart
\markcite{es1991}1991; David et al. \markcite{dsj1993}1993; Allen \&
Fabian \markcite{fa1998}1998; Markevitch \markcite{m1998}1998; Arnaud
\& Evrard \markcite{ae1998}1998). On the other hand, a simple
theoretical model predicts $L_{\rm X} \propto T^2$ on the assumptions
that (1) the internal structures of clusters of different mass are
similar; in particular, the ratio of gas mass to virial mass in the
clusters ($f=M_{\rm gas}/M_{\rm vir}$) is constant;
(2) all clusters identified at some redshift have the same
characteristic density, which scales with the mean density of the
universe (e.g. Kaiser \markcite{k1986}1986; Navarro, Frenk, \& White
\markcite{nfw1995}1995; Eke, Navarro, \& Frenk
\markcite{enf1998}1998). This discrepancy remains one of the most
important problems in clusters of galaxies.

The discrepancy of the $L_{\rm X}-T$ relation is not easily resolved
even if we relax one of these basic assumptions. We just show an
example, where the assumption (1) is relaxed. The X-ray luminosity of
clusters is approximately given by $L_{\rm X} \propto \rho_0^2 R^3
T^{1/2}$, where $\rho_0$ is the characteristic gas density, and $R$ is
the core radius. Thus, the observed relation $L_{\rm X} \propto T^3$
indicates that $\rho_0^2 R^3 \propto T^{5/2}$.  If the gravitational
matter has the same core radius as gas, the baryon mass fraction is
given by $f\propto \rho_0 R^3/RT \propto \rho_0 R^2 T^{-1}$. If we
assume $f\propto T^{\alpha}$, we obtain $\rho_0\propto T^{2-3\alpha}$,
$R\propto T^{-1/2+2\alpha}$, $M_{\rm gas}\propto T^{1/2+3\alpha}$,
$M_{\rm vir}\propto T^{1/2+2\alpha}$, and the characteristic density of
gravitational matter $\rho_{\rm vir} \propto M_{\rm vir}/R^3 \propto
T^{2-4\alpha}$. Assuming that $\rho_{\rm vir}$ is constant in the spirit
of the above assumption (2) (so-called recent-formation approximation),
we should take $\alpha=1/2$.  Thus, this model predicts a correlation of
$\rho_0\propto R \propto T^{1/2}$.  However, such a correlation has not
been found, although many authors have investigated relations among the
physical quantities of clusters (e.g. Edge \& Stewart
\markcite{es1991}1991; Mohr \& Evrard \markcite{me1997}1997; Arnaud \&
Evrard \markcite{ae1998}1998, Mohr, Mathiesen, \& Evrard
\markcite{mme1999}1999). It is to be noted that in the spirit of the
above assumption (1), it is favorable to use core radii when comparing
clusters with different masses, although some previous studies use
isophotal radii instead of core radii in the analysis (e.g. Mohr \&
Evrard \markcite{me1997}1997). Some other studies use a 'virial' radius,
defined as the radius of a sphere of which the mean interior density is
proportional to the critical density of the universe at the observed
redshift of the cluster ($z\sim 0$). However, these radii are derived
from the temperatures of clusters, and are not independent of the
temperatures (e.g. Mohr et al. \markcite{mme1999}1999). Moreover,
$L_{\rm X}$ is mainly determined by structure around core region which
preserves the information of the background universe when the cluster
collapsed (e.g. Navarro, Frenk, \& White \markcite{nfw1997}1997;
Salvador-Sol\'{e}, Solanes, \& Manrique \markcite{ssm1998}1998). Thus,
we adopt the core radius as the characteristic scale of a cluster. Since
most previous works implicitly assumed that clusters form a
one-parameter family, the failure of finding the correlations including
core radii suggests that clusters form a two-parameter family instead.

In this Letter, we reanalyze the observational data of X-ray clusters
and study the relations in detail based on the idea of fundamental
plane.  Originally, the word, 'fundamental plane', represents a
relation among effective radius, surface brightness, and velocity
dispersion of elliptical and S0 galaxies (e.g. Faber et
al. \markcite{fdd1987}1987; Djorgovski \& Davis
\markcite{dd1987}1987). In this study, we apply the notion of the
fundamental plane to X-ray clusters and discuss relations among
$\rho_0$, $R$, and $T$. In \S\ref{sec:data}, results are
presented and in \S\ref{sec:dis}, their implications are
discussed. Throughout the paper we assume $H_0 = 50\;\rm km\; s^{-1}\;
Mpc^{-1}$.

\section{Data}
\label{sec:data}

We use the observational data of the central density, $\rho_0$, core
radius, $R$, and temperature, $T$, of 45 clusters in the catalogue of
Mohr et al. \markcite{mme1999}(1999). We have confirmed that the results
in this section are almost identical to those based on the catalogue of
Jones \& Forman \markcite{jf1984}(1984). Mohr et
al. \markcite{mme1999}(1999) gathered the temperature data of previous
{\em ASCA}, {\em Ginga} and {\em Einstein} observations. On the other
hand, they obtained central densities and core radii using {\em ROSAT}
data; they fitted surface brightness profiles by the conventional
$\beta$ model,
\begin{equation}
  \label{eq:beta}
  \rho_{\rm gas, 1}(r) = \frac{\rho_1}{[1+(r/R_1)^2]^{3\beta/2}} \:,
\end{equation}
where $r$ is the distance from the cluster center, and $\rho_1$, $R_1$,
and $\beta$ are fitting parameters. If an excess in emission (so-called
cooling flow) is seen in the innermost region, Mohr et
al. \markcite{mme1999}(1999) fitted this component by an additional
$\beta$ model,
\begin{equation}
  \label{eq:beta2}
  \rho_{\rm gas, 2}(r) = \frac{\rho_2}{[1+(r/R_2)^2]^{3\beta/2}} \:.
\end{equation}
Since we are interested in global structure of clusters, we use $\rho_1$
and $R_1$ as $\rho_0$ and $R$, respectively. Since Mohr et
al. \markcite{mme1999}(1999) presented only $\rho_2$ for the clusters
with central excess, we calculate $\rho_1$ by
\begin{equation}
 \label{eq:rho1}
 \rho_1 = \left(\frac{I_1 R_2}{I_2 R_1}\right)^{1/2}\rho_2 \:,
\end{equation} 
where $I_1$ and $I_2$ are the central surface brightness corresponding
to the components (\ref{eq:beta}) and (\ref{eq:beta2}), respectively.
Although $R$ and $\beta$ are correlated, each of them was
determined exactly enough for our analysis (see Fig.4 in Mohr et
al. \markcite{mme1999}[1999])

The data plotted in the $(\log \rho_0, \log R, \log T)$ space are fitted
with a plane,
\begin{equation}
  \label{eq:plane}
  A\log{\rho_0} + B\log{R} + C\log{T} + D = 0 \:.
\end{equation}
The result of the least square fitting with equal weight for simplicity
is $A:B:C=1:1.39:-1.29$.  The scatter about the plane is 0.06 dex. This
amounts to a scatter of about 15\%, which is a typical observational
error. We call the plane 'the fundamental plane', hereafter. The ratio
$A:B:C$ is close to $2:3:-2.5$, which is expected when $L_{\rm X}
\propto T^3 \propto \rho_0^2 R^3 T^{1/2}$.  Thus, the observed relation,
$L_{\rm X} \propto T^3$, basically corresponds to a cross section of the
fundamental plane.

In order to study more closely, we investigate further the
distribution of the observational data on the fundamental plane. We
fit the data to another plane,
\begin{equation}
  \label{eq:nplane}
  a\log{\rho_0} + b\log{R} + c\log{T} + d = 0 \:,
\end{equation}
under the constraint,
\begin{equation}
  \label{eq:const}
  Aa+Bb+Cc=0 \:.
\end{equation}
This means that the plane (\ref{eq:nplane}) is perpendicular to the
fundamental plane (\ref{eq:plane}). The result is 
$a:b:c=1:1.18:2.04$. The scatter about the plane is 0.2
dex. We call this plane 'the vertical plane'.
For convenience, two unit vectors in the $(\log \rho_0, \log R, \log
T)$ space are defined by,
\begin{equation}
  \label{eq:e1}
  \mbox{\boldmath $e_1$} = \frac{1}{\sqrt{A^2+B^2+C^2}}(A,B,C) =
  (0.47,0.65,-0.60)\:,
\end{equation}
\begin{equation}
  \label{eq:e2}
  \mbox{\boldmath $e_2$} = \frac{1}{\sqrt{a^2+b^2+c^2}}(a,b,c) =
  (0.39,0.46,0.80)\:.
\end{equation}
Moreover, one of the unit vectors perpendicular to both 
$\mbox{\boldmath $e_1$}$ and
$\mbox{\boldmath $e_2$}$ is defined as $\mbox{\boldmath $e_3$}
=(0.79,-0.61,-0.039)$. The set of
three vectors is one of the bases in the $(\log \rho_0, \log R, \log
T)$ space. Thus, the equations $X=\rho_0^{0.47} R^{0.65} T^{-0.60}$,
$Y=\rho_0^{0.39} R^{0.46} T^{0.80}$, and $Z=\rho_0^{0.79} R^{-0.61}
T^{-0.039}$ are three orthogonal quantities.  Figure 1 shows the cross
section of the fundamental plane viewed from the $Y$ axis.  Figure 2
shows the data on the $(Y,Z)$ plane, i.e., the fundamental plane.  As
can be seen, a clear correlation exists on the plane, that is,
clusters form a band in the $(\log \rho_0, \log R, \log T)$ space.
The major axis of the band is the cross line of the fundamental and
vertical planes, and a vector along the major axis is proportional to
$\mbox{\boldmath $e_3$}$.
We refer to the band as 'fundamental band', hereafter.  Note that the
line determined by the least square method directly from the
three-dimensional data is almost parallel to the vector $\mbox{\boldmath
$e_3$}$. The vector $\mbox{\boldmath $e_3$}$ means that
\begin{equation}
  \label{eq:nR}
  \rho_0 \propto R^{-1.3\pm 0.2} \:,
\end{equation}
\begin{equation}
  \label{eq:TR}
  T \propto R^{0.06\pm 0.1} \propto \rho_0^{-0.05\pm 0.1} \:,
\end{equation}
%
%
Relation (\ref{eq:TR}) indicates that the major axis of the fundamental
band is nearly parallel to the $\log \rho_0 - \log R$ plane, i.e.,
temperature varies very little along the fundamental band. Thus, the
observed relation $L_{\rm X}\propto T^3$ should be the correlation along
the minor axis of the band on the fundamental plane as is explicitly
shown in the next section.

\section{Discussion}
\label{sec:dis}

The results presented in the previous section demonstrate that the
clusters of galaxies are seen to populate a planar distribution in the
global parameter space $(\log \rho_0, \log R, \log T)$. Therefore,
clusters turn out to be a two-parameter family.  The observed relation
$L_{\rm X} \propto T^3$ is a cross section of this 'fundamental
plane'. Moreover, there is a correlation among the data on the
fundamental plane although the dispersion is relatively 
large. This 'fundamental
band' is a newly found correlation between density and radius with a
fixed temperature.

In order to further investigate the relation between physical
quantities and the data distribution in the $(\log \rho_0, \log R,
\log T)$ space, we represent $L_{\rm X}$, $M_{\rm gas}$, $M_{\rm
  vir}$, $f$, and $\rho_{\rm vir}$ by $X$, $Y$, and $Z$, using the
obtained relations
\begin{equation}
  \label{eq:n_0}
  \rho_0 \propto X^{0.47} Y^{0.39} Z^{0.79}\:,
\end{equation}
\begin{equation}
  \label{eq:R}
  R \propto X^{0.65} Y^{0.46} Z^{-0.61}\:,
\end{equation}
\begin{equation}
  \label{eq:T}
  T \propto X^{-0.60} Y^{0.80} Z^{-0.039}\:.
\end{equation}
The results are
\begin{equation}
  \label{eq:lx}
  L_{\rm X} \propto \rho_0^2 R^3 T^{1/2} \propto X^{2.6} Y^{2.6}
  Z^{-0.27}\;,
\end{equation}
\begin{equation}
  \label{eq:gas}
  M_{\rm gas} \propto \rho_0 R^3 \propto X^{2.9} Y^{1.8} Z^{-1.0}\:,
\end{equation}
\begin{equation}
  \label{eq:vir}
  M_{\rm vir} \propto R T \propto X^{0.05} Y^{1.3} Z^{-0.65}\:,
\end{equation}
\begin{equation}
  \label{eq:frac}
  f = M_{\rm gas}/M_{\rm vir} \propto X^{2.4} Y^{0.51} Z^{-0.39}\:,
\end{equation}
\begin{equation}
  \label{eq:rho_vir}
  \rho_{\rm vir} \propto M_{\rm vir} R^{-3} \propto X^{-1.9} Y^{-0.12}
  Z^{1.2}\:.
\end{equation}
%
Exactly speaking, $M_{\rm gas}$ and $M_{\rm vir}$ represent the core
masses rather than the masses of the whole cluster. In relation
(\ref{eq:vir}), we assume that clusters of galaxies are in dynamical
equilibrium. The scatters of $X$, $Y$, and $Z$ are $\Delta \log X =
0.06$, $\Delta \log Y = 0.2$, and $\Delta \log Z = 0.5$,
respectively. Thus, $Z$ is the major axis of the fundamental band and is
the primary parameter of the data distribution. On the other hand,
relation (\ref{eq:T}) indicates that the scatter of $Y$ nearly
corresponds to a variation of $T$ because $\Delta \log T = - 0.60\Delta
\log X + 0.80\Delta \log Y - 0.039\Delta \log Z$. It can be also shown
that a variation of $L_{\rm X}$ is dominated by the scatter of
$Y$. Since $Y$ corresponds the minor axis of the fundamental band, this
means that the $L_{\rm X}-T$ relation is well represented by only the
secondary parameter $Y$, but not by the primary parameter $Z$. To put it
differently, $L_{\rm X} (\propto \rho_0^2 R^3 T^{1/2})$ depends on only
$T$, which is consistent with previous findings. The result reflects the
fact that a combination of $\rho_0$ and $R$ like $\rho_0^2 R^3$ behaves
as a function of $T$ (relation [\ref{eq:nR}]), while $\rho_0$ or $R$
varies almost independently of $T$ (relations [\ref{eq:TR}]). If we
safely ignore the scatter of $X$ and $Z$ in relations (\ref{eq:T}) and
(\ref{eq:lx}), we obtain $T\propto Y^{0.80}$, and $L_{\rm X} \propto
T^{3.3}$. This slope of the $L_{\rm X}-T$ relation approaches the
observed ones, although it is slightly larger.  On the other hand,
$M_{\rm gas}$, $M_{\rm vir}$, and $f$ are not represented by any one of
the parameters $X$, $Y$, and $Z$; both $Y$ and $Z$ contribute to their
variations. Note that $\rho_{\rm vir}$ is mainly governed by $Z$ as
relation (\ref{eq:rho_vir}) shows.

The above analysis raises two questions. The first question is why the
combination like $\rho_0^2 R^3$ behaves as a function of only $T$, or
equivalently why $X$ is nearly constant. In the following arguments, we
assume that the scatter of $X$ is due to observational errors, that is,
$\Delta \log X$ is essentially zero. The behavior of $f$ may be a clue
to the question. Since we allow two parameters, $f$ can be expressed in
terms of any two physical parameters.  For example, if we express $f$
with $M_{\rm vir}$ and $\rho_{\rm vir}$, $f$ turns out to be determined
by $f\propto M_{\rm vir}^{0.4} \rho_{\rm vir}^{-0.1}$.  This means that
the baryon fraction in clusters is an increasing function of $M_{\rm
vir}$. If we adopt the relation of $f\propto M_{\rm vir}^{0.4}$ by hand
and ignore $\rho_{\rm vir}^{-0.1}$ hereafter, we obtain $\rho_0^2
R^{3.2} \propto T^{2.8}$ (relations [\ref{eq:lx}]-[\ref{eq:frac}]),
which is roughly consistent with the shape of the fundamental plane, and
the $L_{\rm X}-T$ relation. Such a relation of $f$ may be realized if
supernovae in the galaxies in the clusters heat the intracluster medium. 
In other words, the behavior of $f$ is likely to originate from the
thermal history of clusters of galaxies.

The second question is why clusters form a two-parameter family. We
think that one natural parameter is $M_{\rm vir}$. As another physically
meaningful parameter, we may choose $\rho_{\rm vir}$. Relation
(\ref{eq:rho_vir}) implies that $\rho_{\rm vir}$ is not constant, which
is inconsistent with the simple theoretical prediction, and that it
varies nearly independent of temperature. Since $\rho_{\rm vir}$ is
supposed to reflect the critical density of the universe when the
cluster, especially around the core region, collapsed, this suggests
that the present day clusters consist of objects with a range of
collapse redshift. In a separate paper, we investigate cosmological
implication of the results presented in this paper (Fujita \& Takahara
\markcite{ft1999}1999).

Finally, we show that the results of this paper reproduce the
size-temperature relation found by Mohr \& Evrard
\markcite{me1997}(1997) and gas mass-temperature relation found by Mohr
et al. \markcite{mme1999}(1999). Since surface brightness profiles of
clusters in the envelope region are given by $I(r) \propto \rho_0^2
T^{1/2} R (r/R)^{-3}$ when $\beta=2/3$, isophotal size, $r=R_{\rm I}$,
has the relation $R_{\rm I} \propto \rho_0^{2/3} R^{4/3}
T^{1/6}$. Eliminating $\rho_0$ by the relation of the fundamental plane,
$\rho_0 R^{1.39}\propto T^{1.29}$, we obtain the relation $R_{\rm
I}\propto R^{0.41} T^{1.03}\propto Y^{1.0}Z^{-0.3}$. This is consistent
with the size-temperature relation $R_{\rm I}\propto T^{0.93}$, although
a coefficient $R^{0.4}$ induces scatter of $\lesssim 30$\% for a given
$T$. The correlation corresponds to a cross section of the fundamental
plane seen slightly inclined from the direction of $Z$ axis. Next, the
consistency with the gas mass-temperature relation is explained as
follows: As in Mohr et al. \markcite{mme1999}(1999), let us define
$R_{\rm vir, m} \propto T^{1/2}$, $M_{\rm vir, m} \propto T^{3/2}$, and
$M_{\rm gas, m}\propto f_{\rm m}\rho_{\rm vir, m} R_{\rm vir, m}^3$,
where $R_{\rm vir, m}$, $M_{\rm vir, m}$, and $M_{\rm gas, m}$ are the
virial radius, the virial mass, and the gas mass of a cluster,
respectively; index m refers to the quantities for $r<R_{\rm vir,
m}$. When $\beta \sim 2/3$, we can show that $f_{\rm m}\propto f$,
because $f_{\rm m}\propto M_{\rm gas, m}/M_{\rm vir, m} \propto \rho_0
R^2 R_{\rm vir, m}/M_{\rm vir, m}\propto \rho_0 R^2 T^{-1}\propto
f$. Since we find $f\propto M_{\rm vir}^{0.4}\propto (RT)^{0.4}$, and
since $\rho_{\rm vir, m}$ is nearly constant by definition, we obtain
the relation $M_{\rm gas, m} \propto R^{0.4} T^{1.9}\propto
Y^{1.7}Z^{0.3}$. This is consistent with the relation $M_{\rm gas,
m}\propto T^{1.98}$ found by Mohr et al. \markcite{mme1999}(1999). Note
that the scatter originated from $R^{0.4}$ is not conspicuous when the
observational data are plotted, because of the steepness of the relation
($\propto T^2$). This correlation also corresponds to a cross section
of the fundamental plane seen from very near to the direction of $Z$
axis.

\acknowledgments

This work
was supported in part by the JSPS Research Fellowship for Young
Scientists.

\newpage 

\begin{figure}
\centering \epsfig{figure=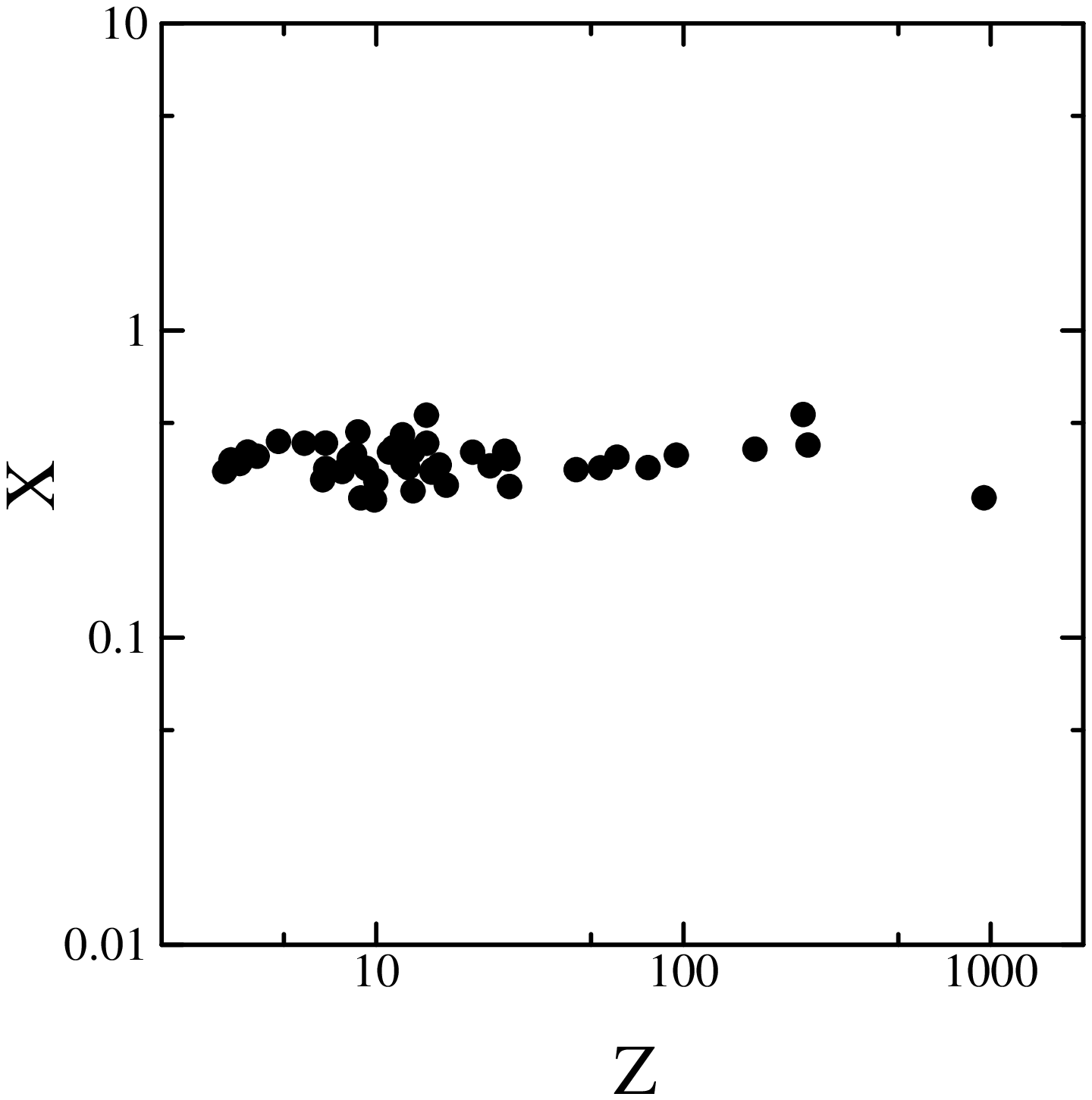, width=8cm} 
\caption{The observational data projected on the $Z-X$ plane.
} 
\end{figure}

\begin{figure}
\centering \epsfig{figure=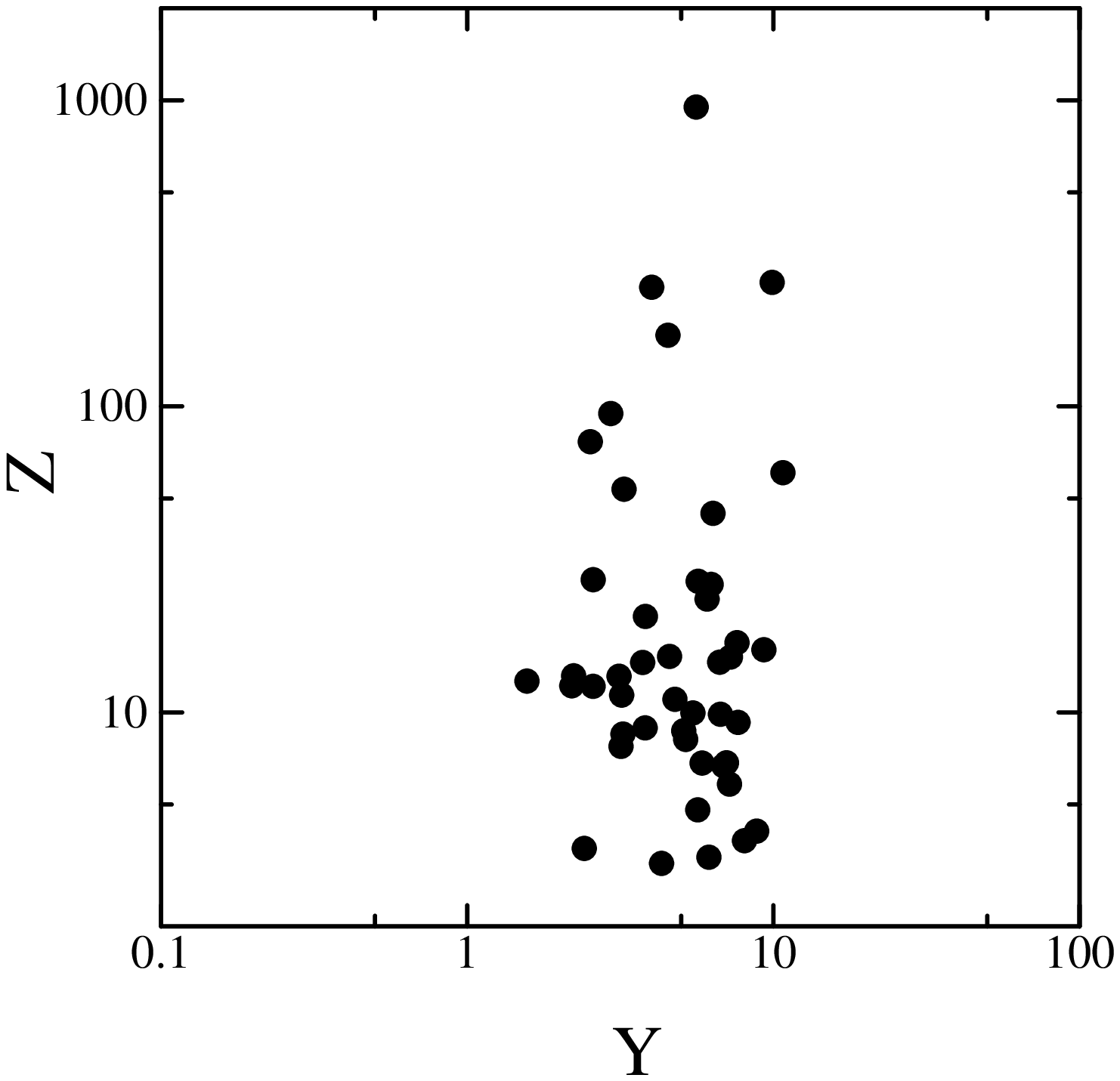, width=8cm} 
\caption{The observational data projected on the $Y-Z$ plane.
} 
\end{figure}

%
%

\end{document}